# Recommendations on Statistical Randomness Test Batteries for Cryptographic Purposes

ELENA ALMARAZ LUENGO*, Group of Analysis, Security and Systems (GASS), Department of Statistic and Operational Research, Faculty of Mathematical Science, Universidad Complutense de Madrid (UCM)

LUIS JAVIER GARCÍA VILLALBA†, Group of Analysis, Security and Systems (GASS), Department of Software Engineering and Artificial Intelligence (DISIA), Faculty of Computer Science and Engineering, Office 431, Universidad Complutense de Madrid (UCM)

Security in different applications is closely related to the goodness of the sequences generated for such purposes. Not only in Cryptography but also in other areas, it is necessary to obtain long sequences of random numbers or that, at least, behave as such. To decide whether the generator used produces sequences that are random, unpredictable and independent, statistical checks are needed. Different batteries of hypothesis tests have been proposed for this purpose.

In this work, a survey of the main test batteries is presented, indicating their pros and cons, giving some guidelines for their use and presenting some practical examples.

## 1 INTRODUCTION

In recent years, due to an increase in electronic transactions and the dissemination of sensitive information via networks, ensuring the security of the involved information has become quintessential. Therefore, there has been an increase in cryptographic studies with an aim to obtain better encryption systems that, as well, require unpredictable values (random numbers) which are essential to the efficacy and success of any cryptographic tool. To precisely guarantee this security in data processing and the confidentiality of the information being transmitted, the numbers used in the generation of the different keys involved must exhibit certain properties: they must pass certain statistical tests that

Authors' addresses: Elena Almaraz Luengo, ealmaraz@ucm.es, Group of Analysis, Security and Systems (GASS), Department of Statistic and Operational Research, Faculty of Mathematical Science, Universidad Complutense de Madrid (UCM), Plaza de Ciencias 3, Ciudad Universitaria, Madrid, Spain, 28040; Luis Javier García Villalba, javiergv@fdi.ucm.es, Group of Analysis, Security and Systems (GASS), Department of Software Engineering and Artificial Intelligence (DISIA), Faculty of Computer Science and Engineering, Office 431, Universidad Complutense de Madrid (UCM), Calle Profesor José García Santesmases 9, Ciudad Universitaria, Madrid, Spain, 28040.

ensure (with a certain level of confidence) the randomness and uniformity of the sequences used, thus rendering the sequences irreproducible. Furthermore, any given series of generated values cannot be obtained by any function or mathematical model. The verification of the statistical tests is thus crucial to the development of efficacious sequences and to avoid predictable mathematical models that may potentially breach security systems.

Random numbers are the fundamental premise of simulation. It would be desirable to obtain pure random numbers that verify the conditions of independence and uniformity and which are not reproducible. In other words, pure random numbers allow the accomplishment of safe operations is the target. Commonly, generators are used to provide sequences that seem to be independent realizations of a Uniform random variable in the interval (0,1) ($U(0, 1)$), and we say "apparently" because, in general, they tend to use mathematical algorithms that generate sequences that, although generated with a particular algorithm, verify the desired conditions abovementioned, namely, pseudo-random sequences. According to this, we can define two main classes of random numbers (this is a very common classification in literature. See section 2.2. of [50] or section 2.3. of [59]): "true" random numbers and pseudo-random numbers. There is one additional type, also defined in literature, known as quasi-random numbers.

*Definition 1.1 (True Random Numbers).* True random numbers are based on physical sources (entropy sources) which are sources of intrinsic natural randomness along with some processing function (see [52]). Therefore, they do not need an initial sequence (also known as "seed") and are not expected to show periodic patterns or correlations between the obtained data.

Other advantages of true random numbers are the high level of security they provide in some applications (such as key generation) and that since there is no underlying algorithm that generates them, we should not worry about them being reproduced. On the contrary, some disadvantages are that their generation is time-consuming and computationally inefficient (hence, the speed of generators of this type of numbers cannot match the demand of bitrate that is often required), their installation and performance are troublesome and its generation is difficult. An interesting description of this type can be seen in [52].

*Definition 1.2 (Pseudo-random Numbers).* Pseudo-random numbers constitute a series of values which, although are deterministically generated (as they are constructed through generation algorithms and an initial seed), have the appearance of being values that come from independent realizations of a uniform random variable [49].

The way pseudo-random numbers are generated makes them reproducible.

*Definition 1.3 (Quasi-random Numbers).* Quasi-random numbers are obtained through specific algorithms and the obtained sequences are distributed in a uniform way in the square or in the cube unit. For a (finite) quasi-random sequence the objective is that it fills a unit hypercube as uniformly as possible [23].

As the dimension increases, the main disadvantage of quasi-random numbers is that there are not specific hypothesis tests to evaluate the goodness of the obtained sequences [39].

In Cryptography, the most common generators used are random number generators (RNGs) and pseudo-random number generators (PRNGs), both of which produce bit sequences (0-1). For the purposes of Cryptography, these obtained numbers must be unpredictable. This concept is intimately related to the absence of correlations between the obtained numbers and their independence. The problem of randomness testing is becoming increasingly crucial, particularly because of the need of effective security of communications [51] due to today's widespread use of public communications, e-commerce (key management, electronic signatures), etc.

In the case of RNGs, a non-deterministic seed (entropy source) and a generating function are necessary to produce randomness. The source of the entropy generally comes from some physical phenomenon like the noise of an electrical circuit, hardware sources, etc. For more details you can see [64] or [52].

This type of generator can potentially become problematic if the source of entropy is predictable. In order for this not to affect the sequence obtained in terms of its fundamental condition of unpredictability, it would be required to combine outputs obtained with different sources. However, this solution is not optimal since the obtained sequences can be statistically inadequate. They could not pass the randomness or uniformity tests. Contrastingly, from a computational point of view, the generation of large sequences is ultimately time-consuming. This means that there exists a tendency to use PRNGs for their immanently better production of statistical and computational results.

As previously commented, in the case of the PRNGs, if the seed is known, given that the structure of the generator is fixed, the complete sequence could be determined, which is why special care must be taken to keep the seed secret. The seed can be generated by using a RNG.

There is a variety of studies on the generation of random numbers in the Computer Science area, Information Theory and Cryptography. Some interesting references are [18] or [33] in whose works aspects related to the generation of random numbers and information theory appear. In [27] a theoretical framework is established for the generation of random sequences from the perspective of information theory.

In [60], the authors perform a detailed study on the generation of Gaussian random numbers. In [28], the authors address the uniform random number generation without information leakage and address the secure uniform random number generation with partial information leakage.

In [46], a new methodology to generate pseudo-random numbers is presented by using generative adversarial networks. To this effect, the authors validate their result by using NIST tests suite. Previously, in 1994, a method for generating pseudo-random numbers was presented (digital inversive method) [17].

On the other hand, works in which the properties of generators are studied in detail can also be found. For instance, in [21] properties of nonlinear filter of m-sequences (an important type of pseudo-random sequence generator) are studied. In [26], the authors study the power generator of pseudo-random numbers in the particular case of Blum-Blum-Shub generator. In [22], the authors obtain a class of keystream generators with a large linear complexity.

There are also several works in which weaknesses of existing generators are revealed as in [3], where the author shows that the Marsaglia and Zaman's generator produces the successive digits of a rational b-adic number and, in this work, an efficient prediction algorithm for the sequences generated by this generator is obtained. In addition, there are other works which suggest improvements in the sequences obtained such as in [35], where a post-processing function to reduce or eliminate statistical weaknesses of physical random number generators is used; or in [1], where a compression method (using pseudo-chaotic systems) is applied to True Random Bit Generators.

As showcased in previous lines, the necessity of "good sequences" is a fact and a point of great interest. As a result, it is necessary to verify if the sequences meet the desired requirements. From the statistical point of view, certain tests must be passed to ensure that the sequences obtained meet the desired conditions.

This article surveys the principal statistic batteries existing in the literature, analyzes them and gives some recommendations about their use. The rest of the paper is organized as follows: in Section II the principal concepts about hypothesis testing are given, in Section III, the principal batteries are explained and some examples are given, in Section IV other hypothesis tests that are not included in the batteries described in the previous section are described. In section V some recommendations about the use of the batteries are given. Finally, in Section V the conclusions are given.

## 2   HYPOTHESIS TESTING

In order to verify if the sequence obtained with a generator can be considered as a "true random sequence", different hypothesis tests can be applied, given a specific level of significance to decide on this question. Randomness is a statistical property, that is, the sequence will be characterized in probabilistic terms and its properties will be compared with those expected in the case of truly random sequences.

In the following, the concept of hypothesis testing will be defined from a statistical perspective.

The probability function of a random variable (r.v.) $X$, $f_X(x,\theta)$ depends on a parameter $\theta$ that takes values in a parametric space $\Theta$, in such a way that for each value $\theta \in \Theta$, the function $f_X(x,\theta)$ is different. A statistical hypothesis about the parameter is a guess about the specific values it may take.

The establishment of a hypothesis on the parameter implies dividing the space $\Theta$ into two separate parts, $\Theta_0$ (formed by the set of values $\theta$ that satisfy the hypothesis) and $\Theta_1$ (formed by the set of values $\theta$ that do not satisfy the hypothesis) so that $\Theta = \Theta_0 \cup \Theta_1$.

The hypothesis to be contrasted is called Null Hypothesis $H_0$ and the other, Alternative Hypothesis $H_1$. The problem that arises in a hypothesis contrast is to determine which of the two hypotheses can be accepted.

A hypothesis contrast is a rule of decision by which one hypothesis or another is chosen in light of the information obtained through a sample taken from the population under study.

The solution given to the problem of contrasting the two hypotheses implies the possibility of succeeding or failing in the choice because of not knowing with certainty which is the true one. The situation is reflected in Table 1 (for more details see [8]):

| TRUE SITUATION | CONCLUSION | |
|---|---|---|
| | Not reject $H_0$ | Reject $H_0$ (Accept $H_1$) |
| $H_0$ is true | No error | Type I error |
| $H_1$ is true | Type II error | No error |

Table 1.  Types of errors in hypothesis testing.

- If $H_0$ is true and it is not rejected, the decision is correct.
- If $H_0$ is true and it is rejected, the decision is incorrect and this error is called Type I Error.
- If $H_1$ is true and it is not rejected, the decision is correct.
- If $H_1$ is true and it is rejected, the decision is incorrect and this error is called Type II Error.

The above situations are unknown and uncontrolled to some extent. However, a certain degree of control can be established over them by knowing the probabilities of committing the above errors. The probability of Error Type I is defined as $\alpha$ and it is also known as the level of significance of the contrast, generally established in values of 0.01, 0.05 or 0.10. The probability of Error Type II is defined as $\beta$. In practice, it is usually used $1 - \beta$ or contrast power. The level of significance and contrast power are not independent.

In practice, the concept of $p - value$ is often used. The $p - value$ is defined as the probability of obtaining results of the test, at least, as extreme as the results actually observed. If the condition of $p - value < \alpha$ is satisfied, then $H_0$ will eventually be rejected.

It is interesting to note that we do not always work with an underlying distribution $f_X(x,\theta)$ of the population but this does not mean that we cannot carry out a hypothesis analysis, that is, non-parametric techniques have been developed. In other words, no hypotheses appear about a certain parameter $\theta$ of a population, but the process is based on a statistician who will not refer to the parameter. It is common to use certain sample information such as the frequency with which the data appear in the sample, its position, etc. In this type we can highlight goodness-of-fit contrasts or randomness contrasts.

For cryptographic purposes, hypothesis contrasting should be used to determine whether the sequence of numbers obtained by a generator satisfy the uniformity hypothesis (in the sequence of random or pseudo-random bits obtained, the expected number of zeros and ones is the same, i.e., if $n$ is the length of the sequence, the expected number of zeros (ones) is $n/2$), scalability (if a sequence is random, any substring obtained from the first must also be random) and consistency (the behavior of the generator must be studied for different seeds, and the result of the sequence obtained in terms of its randomness must not depend on the chosen seed).

The most common tests used to decide on the uniformity of the sample are the tests $\chi^2$ and Kolmogorov-Smirnov (K-S).

The steps of the $\chi^2$ test are [49]:

(1) Organize the $n$ pseudo random numbers in $k$ disjoint classes $I_i$, $1 \leq i \leq k$. Count the quantity of $u_i \in I_i$, denote this number by $f_i$. So $f_1 + \ldots + f_k = n$ is the size of the sample. Under $H_0$, $P(I_i) = p_i$ where $p_1 + \ldots + p_k = 1$, because the sample is composed by independent items, $f_i \sim B(n, p_i)$, where $B(n, p_i)$ represents a Binomial distribution with parameters $n$ and $p_i$.

(2) Consider the variable $V = \sum_{i=1}^{k} \frac{(f_i - np_i)^2}{np_i}$. The asymptotic distribution of $V$ is chi-square with $k-1$ degrees of freedom ($\chi^2_{k-1}$) when $n \to \infty$.

(3) Select $\alpha \in (0,1)$. Get $\chi^2_{k-1,\alpha}$ from the quantiles table o the $\chi^2$ distribution.

(4) If $V > \chi^2_{k-1,\alpha}$ reject $H_0$. If $V \leq \chi^2_{k-1,\alpha}$ there is no evidence for rejecting $H_0$ at level $\alpha$.

K-S test consist of the following steps [49]:

(1) Arrange the pseudo-random numbers $u_i$ from order from low to high. Let $\left\{ y_i \right\}$ be the resulting sequence.

(2) Build the empirical distribution function $F_n(y_i) = i/n$, $i = 1, \ldots, n$.

(3) Compute de KS statistic: $D_n = \max_{1 \leq i \leq n} |F_n(y_i) - y_i|$, $0 \leq y_i \leq 1$.

(4) Select $\alpha \in (0,1)$ that represents the probability of rejecting $H_0$. Get $D_{n,\alpha}$ from the table of KS quantiles.

(5) If $D_n \leq D_{n,\alpha}$ there is no evidence for rejecting $H_0$ at level $\alpha$. If $D_n > D_{n,\alpha}$ reject $H_0$.

The principal differences between them are:

- K-S is easier to apply than $\chi^2$.
- K-S is not affected by the regroupings.
- K-S is used for small samples.
- The power of K-S test is higher than $\chi^2$ test.
- Test $\chi^2$ can be adapted in an easy way for the case of unknown population parameters.
- Test $\chi^2$ can be used in case of discrete and continuous distributions.

These tests are widely used in various batteries that will be exposed in the next section. Its use is direct and indirect, i.e., in some of them, they appear as elements of the battery and, in other cases, the distributions and tests are used as tools for other tests of the batteries. To contrast the hypothesis of randomness there are also several approaches. One of which is the runs test, the best known, which will be explained later.

## 3   BATTERIES OF TESTS USED FOR HYPOTHESIS TESTING IN CRYPTOGRAPHY

The sequences of bits obtained by the generators are expected to pass the hypothesis tests (i.e. there is no evidence to reject the null hypothesis for the selected $\alpha$ level) of interest to contrast the desired cryptographical properties previously commented on.

In literature, there are several batteries that have been traditionally used in cryptography. Many authors have studied some particular examples or have given a general overview about them, for example, in [34], some of the most popular tests are described. In [51], the author explains some statistical tests depending on their focus, which is the aspect they analyze. In section 6 of [58], or in section 2 of [10] the most known batteries are mentioned.

In the following subsections, our objective is to explain in general terms the tests that compose the different batteries and to indicate the pros and cons of each one.

### 3.1   Knuth's battery

The first test battery used in Cryptography was introduced by Knuth [34]. Its use is currently set apart and other options (other batteries) are preferred, for example, for its computational implementation, its facility in their uses, etc. It includes 12 statistical tests that are described briefly in the following lines.

- Frequency test: it evaluates whether the elements of a sequence are uniformly distributed. K-S test is used for this purpose.

- Serial test: it is tested if each of the elements of the sequence happens with the same frequency and additionally if the pairs of possible results obtained from the sequence are independent and uniformly distributed. The test can be generalized for lists of three elements, four elements, etc.

- Gap test: for each element in the sequence the amount of elements different from that element is counted before it happens again, the process is repeated for all its occurrences, and performs a $\chi^2$ test on the lengths of the gaps.

- Poker test: this test considers $n$ groups of 5 consecutive integers and classifies them according to the following seven patterns: the five integers that appear are all different ($abcde$), an integer is repeated twice and three different ones appear once each ($aabcd$), an integer is repeated twice, a different one is repeated twice and a third different integer appears only once ($aabbc$), an integer is repeated three times and two different ones appear once each ($aaabc$), an integer is repeated three times and a different one is repeated twice ($aaabb$), the same integer is repeated four times and a different one appears once ($aaaab$), the same integer is repeated five times ($aaaaa$). Under the hypothesis of randomness and adjustment to a $U(0, 1)$, the probabilities of these modalities can be computed. Classes $aaaaa$ and $aaaab$ are often grouped together when applying the $\chi^2$ test.

- Coupon collector's test: this test examines the amount of data needed to obtain at least one of each of the possible values. The lengths obtained are examined with the $\chi^2$ test.

- Permutation test: this test considers blocks of $k$ elements given a sequence of size $n$. Each of the blocks can have any of the $k!$ possible arrangements of these $k$ values. If the sequence were random, each possible arrangements would occur with equal probability: $1/k!$. The test consists of observing many blocks and compare the observed frequencies of each possible order with the expected ones by applying a $\chi^2$ test.

- Runs test: the test is based on the length of the ascending or descending runs, understood as ascending run when each item in the series is greater than the previous one, and a descending run when each item in the series is less than the previous one. The number $R$ of runs (+ if $u_i < u_{i+1}$, − if $u_i > u_{i+1}$) is calculated, the asymptotic distribution of $R$ is $N(\mu = \frac{1}{3}(2n-1), \sigma = \frac{1}{90}(16n-29))$, where $N(\mu, \sigma)$ represents the normal distribution with mean $\mu$ and variance $\sigma^2$, consider $Z = \frac{R-\mu}{\sigma}$ whose asymptotic distribution is $N(0, 1)$, then a $\alpha$ level is selected, the related quantile $z_{\alpha/2}$ such that $P\left(Z \geq z_{\alpha/2}\right) = \alpha/2$ is taken and the randomness is accepted if and only if $|Z| < z_{\alpha/2}$.

- Maximum of $t$ test: the sequence is divided into subsequences of equal length, the maximum value of each subsequence is selected and a K-S test is applied to the selected values.

- Collision test: this test is used in cases where the amount of data is significantly less than the number of categories to which they may belong, so that the probability that two or more data belong to the same category is low. Collision is defined as the case in which two data belong to the same category.

- Birthday Spacings test: $m$ data from the sequence being tested are selected, which can be between 1 and $n$, they are treated as if they were $m$ birthday in a year of $n$ days, then the $m$ days are ordered and the gaps between each pair are calculated and these gaps are grouped by size. The particular description will be given in the description of Diehard tests battery.

- Serial Correlation test: the serial correlation statistic is computed. This statistic measures how much one element of the sequence depends on the previous one. If the sequence were random, the serial correlation values would have values very close to zero.

- Subsequences test: this test is applied in cases where more than one random number is required at a time.

Originally Knuth's battery was developed for simulation applications the same as the next battery we will explain, Marsaglia's battery.

## 3.2 Diehard tests battery

In 1995, George Marsaglia [40] published a CD-ROM of random numbers along with a set of tests to determine if a sequence of numbers could be considered as a random sequence. He called this battery "Diehard tests". If a PRNG passes the tests on this battery, then it can be used in many quality scientific studies. In this battery, 16 tests are included which are briefly described below (for more details you can see the official site [40] or section 3 of [24] in which a summary of the tests is given).

- Birthday spacing test: in this test $m$ birthdays of a year of $n$ days are selected, then a list of the time between two events (two consecutive birthdays) is made. Let $j$ be the number of values that appear more than once in that list, then the asymptotic distribution of $j$ is Poisson of parameter $\lambda = m^3/(4n)$, $P\left(m^3/(4n)\right)$ (in practice that approximation is valid when $n \geq 2^{24}$ and $m = 2^9$). This test recommends $n > 2^{18}$ and in practice takes $n = 2^{24}$ and $m = 2^9$ so the distribution to work with in this case is $P$ $(\lambda = 2)$. Samples of size 500 are taken, and a $\chi^2$ test of goodness of fit is done which provides a $p - value$. The first test uses the 1-24 bits (counting from the left) of integers in the file. Then the file closes and reopens. The second one uses the bits 2-25, the third uses 3-26 bits and so on up till bits 9-32. Each set of bits provides a $p - value$, and the nine $p - values$ forms a sample to which a K-S test is applied. This is one of the most difficult test to pass and if a sequence passes this test, it is very common that that sequence passes most of the rest tests of this battery.

- Overlapping 5-permutation test (OPERM5): a sequence of one million 32-bit random integers is being studied. A set of 5 consecutive integers is selected and the 120 possible permutations between the 5 selected numbers must appear with equal probability along the rest of the analyzed series.

- Binary rank tests (for 31x31, 32x32 and 6x8 matrices): the aim of these tests is to check whether the rank of the obtained data matrices is consistent with the expected rank for matrices whose entries are random.

- Bitstream test: the data set is considered to be a bit stream $a = \{a_i\}$ and a two-letter alphabet $(0 - 1)$ is considered, so the $a$ flow can be seen as a succession of words of 20 overlapping letters (the first word would be $a_1 a_2 \ldots a_{20}$, the second would be $a_2 a_3 \ldots a_{21}$ and so on). This test counts the number of words of 20 letters (20 bits) that are missing in a string of $2^{21}$ words of 20 overlapping letters and compares it with the expected number if the sequence were random. There are $2^{20}$ possible words of 20 letters. If the sequence were random with $2^{21} + 19$ bits, the number of missing words $f$ would be distributed according to a random variable $N(\mu = 141.909, \sigma = 428)$, then, the number obtained in the sequence under study is compared with the quantile of the standard Normal distribution $Z = \frac{f - \mu}{\sigma}$. The test must be repeated 20 times.

- Tests overlapping-pairs-sparse-occupancy (OPSO), overlapping-quadruples-sparse-occupancy (OQSO) and DNA: the OPSO test considers words of 2 letters chosen from an alphabet of 1024 letters. Each letter is determined by a dozen bits of a 32-bit integer in the sequence under analysis. The test generates $2^{21}$ (overlapping) words of 2 letters (from $2^{21} + 1$ "keystrokes") and counts the number of words of 2 letters that does not appear in the sequence under analysis. If the sequence were random, this amount should be distributed according to a random variable $N(\mu = 141909, \sigma = 290)$, therefore, this test compares the number obtained in the concrete sample with what theoretically should be obtained through the normal random variable. This test takes 32 bits at a time from the test file and uses a designated set of ten consecutive bits. Then it restarts the file for the next 10 designated bits, and so on. The OQSO test is similar to the previous one but considering 4-letter words of a 32-letter alphabet where each letter is determined by a designated string of 5 consecutive bits of the test file, whose elements are assumed as random 32-bit integers. The DNA test considers a 4-letter alphabet $(C, G, A, T)$, determined by two designated bits in the sequence of random integers that is being tested. We then consider 10-letter words and test whether the number of missing words is compatible with the theoretical value that should be there, which is distributed according to a $N(\mu = 141909, \sigma = 339)$.

- The count-the-1's test on a stream of bytes: this test considers the file of numbers as a sequence of bytes (4 per 32-bit integer). Each byte can contain an amount of ones that varies in $\{0, \ldots, 8\}$ with probabilities 1/256, 8/256, 28/256, 56/256, 70/256, 56/256, 28/256, 8/256 and 1/256 respectively. The byte set provides an overlapping word of 5 letters, each letter can take values in the set $\{A, B, C, D, E\}$. The letters are determined by the number of ones in that byte (so, the letter to be taken depends on the number of ones in the sequence): numbers $0, 1, 2$ correspond to $A$, 3 to $B$, 4 to $C$, 5 to $D$ and finally, $6, 7, 8$ to $E$. The number of overlapping words of 5 letters is $5^5$. From a string of 256000 words of 5 letters (overlapped), counts are made in the frequencies of each word. The quadratic form in the weak inverse of the covariance matrix of the cell counts provides a $\chi^2$ test. The test returns two $p - values$ for both 5 and 4-letter cell counts.

- The count-the-1's test for specific bytes: this test studies the randomness of the sequence of 5-letter random words that are overlapping. It is similar to the previous one, in this case, the test is applied 25 times, first using the first byte, then the second byte and so on, and the corresponding $p - values$ of $\chi^2$ test are found.

- The parking lot test: in this test a car is parked (it is a circle with radius 1) in a square of size 100x100. Then a second car is parked, then the third and so on. If a crash occurs the process for that particular car is replicated from the beginning choosing different random places for parking. Let $n$ be the number of attempts and $m$ the number of successfully parked, in practice it is chosen $n = 12000$, simulation shows that $m$ distributes (asymptotically) as a $N(\mu = 3523, \sigma = 21.9)$. At the end a K-S test for 10 obtained $p - values$ is performed to check if they come from a $U(0, 1)$ distribution.

- Minimum distance test: a square is taken again but in this case of side 10000 and 8000 points are chosen randomly in it. Let $d$ be the minimum distance between the $\frac{n(n-1)}{2}$ pairs of points. If the points were evenly distributed, $d^2$ would be distributed according to an exponential random variable with parameter $\lambda = 1/0.995$ ($\exp\{\lambda = 1/0.995\}$), then $1 - \exp\{-d^2/0.995\}$ should be distributed according to a $U(0, 1)$ random variable and a K-S test is applied to the resulting 100 values in order to check the uniformity for random points in the square.

- 3D spheres test: the aim of the test is to evaluate the randomness of triplets of sequential random numbers of uniform distribution. They are 4000 random points chosen on a 1000 side cube. At each point, we center a sphere sufficiently large to reach the next nearest point. The volume of the smallest sphere is distributed approximately according to an $\exp\{\lambda = 3/120\pi\}$ random variable, so the radius of the cube is distributed according to an $\exp\{\lambda = 1/30\}$ random variable, the average $\mu = 30$ is obtained through extensive simulation. This test generates 4000 spheres 20 times. Each min radius cubed results in a uniform variable through $1 - \exp\{-r^3/30\}$, then a K-S test is applied to the 20 $p - values$.

- The squeeze test: starting with $k = 2^{31}$, multiply by floating point values on $[0, 1)$ until $k = 1$. Let $t$ be the number of iterations that is necessary to reduce $k$ to 1. We repeat this test 100000 times, and we count the number of times in which $t < 7$ and $t \geq 47$ and compare them to an expected exponential to get the $p - value$.

- The overlapping sums test: the sequence of integers is floated to get a sequence $U_i$ of values from a $U(0, 1)$ distribution. Then the sums $S(j) = \sum_{i=j}^{100+j-1} U_j$ are defined. The $S(j)$ have an approximate Normal distribution, they are typified to obtain a set of independent values from a distribution of $N(0, 1)$ which are transformed into uniform variables for a K-S test. The $p - values$ from ten K-S tests are given still another K-S test.

- The runs test: the number of runs (increasing and decreasing) is counted in the sequence of values to be tested. Runs are counted for sequences of length 10000. The procedure is similar to the description previously explained, it is done 10 times, and then it is repeated.

- The craps test: it is considered a game consisting of 200000 dice games, the number of wins $n$ and the number of throws needed to finish each game $l$ are counted. $n \sim$ (asymptotically) $N(\mu = 200000p, \sigma = \sqrt{200000p(1 - p)})$, with $p = 244/495$. $l \in [1, \infty)$, but counts for all players greater than 21 are grouped with 21. A $\chi^2$ test is made on the number of-throws. Each 32-bit integer taken from the file to be evaluated provides the value for the throw of a die by normalizing it to an interval $[0,1)$ and then multiplying it by 6 and adding 1 to the integer part of the result.

There are some disadvantages in this battery. The main one of them is that the parameters that appear in the tests are set by the software and cannot be modified like the allowed sample size. It is required that the sequences to be tested are in a binary file in the form of 32-bit integers (exactly). In addition, there are also difficulties regarding the technical handling of the same, as for example in the data input [36].

### 3.3 Dieharder battery

In order to solve the disadvantages of the previous battery, Brown et. al. [6] introduced the Dieharder battery which consists of 26 tests, including Diehard battery tests, K-S test, runs test and the following are incorporated:

- Greatest Common Divisor Marsaglia-Tsang: this test is based on the calculus of the Greatest Common Divisor (GCD) of two numbers. Two random 32-bit positive integers $u$ and $v$ ($u,v \in [1, 2^s]$) are chosen and Euclid's algorithm is applied to them. As a result, three elements are obtained: (1) the number of iterations $k$ that have been done to find GCD, (2) a variable length sequence of partial quotients and (3) the GCD of $u$ and $v$. If $k$ and GCD are considered as random variables, these are independent and identically distributed. For $s$ big, the mass function of GCD is approximately $P\left[GCD\left(u,v\right)=j\right]=\frac{6}{\pi^2 j^2}$ and the distribution of $k$ is approximately Binomial with parameters $n=50$ and $p=0.376$ (for a complete description of this test see [41]). A standard goodness-of-fit test for $\chi^2$ distributions is applied to both GCD and $k$. Deviations from randomness suggest nonconformity between empirical and theoretical distributions of $k$ and GCD.

- Generalized minimum distance: Diehard Minimum distance test generalized to higher dimensions, see [19].

- Lagged sum test: this test calculates the mean of the sequence with $n$ lags and compares it with the theoretical value.

- Permutations test: this test is similar to the Non-Overlapping Serial Test but with a different mapping of the input uniforms to the sample statistics. This test counts order permutations of random numbers, taken $n$ at a time. There are $n!$ permutations equally probable and the samples are independent. In this way a simple $\chi^2$ test on the count vector with $n!-1$ degrees of freedom can be applied.

- Monobit: this test studies the number of zeros and ones in the sequence under study. If the sequence is random, the amount is expected to be approximately the same. The study can be done in terms of proportions, the proportion of zeros and ones is expected to be approximately the same. This test is also known as a frequency test.

- Generalized Serial test: this test attempts to verify whether the sequences of successive numbers are equally distributed independently or not. This is a generalization of the serial test, see [25], [32] or [34].

- Bit distribution test: this test performs the accumulation of the frequencies of all tuples of bits of size $n$ (selected without overlap) in a list of random integers and compares the empirical distribution of the sample with the theoretical (binomial) distribution by using a $\chi^2$ test.

These tests are open source and can be used in R interface in Linux or Unix operating systems [6].

Further studies have been carried out to speed up the execution of the tests of this battery, see for example [62], where a solution is proposed to accelerate the statistical tests based on reconfigurable hardware, taking advantage of the parallelization of tasks and high frequencies.

In order to apply this battery we should go to the *dieharder* website:

$https : //webhome.phy.duke.edu/ r\"ob/General/dieharder.php$ [6].

In this webpage the Diehard battery can be downloaded. There are different versions and one can find the information about its installation and use. The files have extension *.rpm* and *.tgz*. Originally the files are prepared to be used under Linux or Unix, but there is the possibility to work under Windows operating system through the R interface.

Next we will show some examples of the application of this battery by using R interface. For this purpose we need to work with the library *RDieharder* [16]. The available tests can be seen in Figure 1. Moreover, there are different and known generators that can be used in R, see Figure 2.

In Example 1, see Figure 3, we use 100 samples of numbers generated by the generator *mt19937* (Mersenne Twister 19937 generator) and seed 12345. We apply the test number 15. The $p-value$ that is calculated is $p=0.8347$, which is higher than 0.01, so the sequence clearly passes the test.

In Example 2, see Figure 4, we use 100000 samples of numbers generated by the generator *randu* (particular linear congruential generator with parameters $m=2^{31}$, $a=65539$ and $b=0$) and seed 12345. We apply the test number 15. In this case $p-value < 2.2e^{-16}$, so the sequence clearly does not pass the test.

| | names | id |
|---|---|---|
| > dieharderTests() | | |
| 1 | diehard_birthdays | 0 |
| 2 | diehard_operm5 | 1 |
| 3 | diehard_rank_32x32 | 2 |
| 4 | diehard_rank_6x8 | 3 |
| 5 | diehard_bitstream | 4 |
| 6 | diehard_opso | 5 |
| 7 | diehard_oqso | 6 |
| 8 | diehard_dna | 7 |
| 9 | diehard_count_1s_stream | 8 |
| 10 | diehard_count_1s_byte | 9 |
| 11 | diehard_parking_lot | 10 |
| 12 | diehard_2dsphere | 11 |
| 13 | diehard_3dsphere | 12 |
| 14 | diehard_squeeze | 13 |
| 15 | diehard_sums | 14 |
| 16 | diehard_runs | 15 |
| 17 | diehard_craps | 16 |
| 18 | marsaglia_tsang_gcd | 17 |
| 19 | sts_monobit | 100 |
| 20 | sts_runs | 101 |
| 21 | sts_serial | 102 |
| 22 | rgb_bitdist | 200 |
| 23 | rgb_minimum_distance | 201 |
| 24 | rgb_permutations | 202 |
| 25 | rgb_lagged_sum | 203 |
| 26 | rgb_kstest_test | 204 |
| 27 | dab_bytedistrib | 205 |
| 28 | dab_dct | 206 |
| 29 | dab_filltree | 207 |
| 30 | dab_filltree2 | 208 |
| 31 | dab_monobit2 | 209 |
| 32 | dab_birthdays1 | 210 |
| 33 | dab_opso2 | 211 |

Fig. 1. Tests in Dieharder battery available in *RDieharder*.

### 3.4   NIST battery

The US National Institute of Standards and Technology (NIST) in 2001 [52] developed a test suite called NIST battery, consisting of 15 randomness tests. The NIST battery is widely used mainly for quality certifications. As with the Dieharder battery, test parameters are fixed and this reduces its flexibility. NIST battery was developed to detect non randomness for cryptographic applications. Next, we will explain briefly the tests included in this battery, but for a complete guide it is recommended reading the official documentation [52] (also, there is a lot of bibliography in which some of these tests are explained, for example in [57]):

```
> dieharderGenerators()
```

| | names | id | | names | id | | names | id |
|---|---|---|---|---|---|---|---|---|
| 1 | borosh13 | 0 | 28 | random256-glibc2 | 27 | 54 | taus2 | 53 |
| 2 | cmrg | 1 | 29 | random256-libc5 | 28 | 55 | taus113 | 54 |
| 3 | coveyou | 2 | 30 | random32-bsd | 29 | 56 | transputer | 55 |
| 4 | fishman18 | 3 | 31 | random32-glibc2 | 30 | 57 | tt800 | 56 |
| 5 | fishman20 | 4 | 32 | random32-libc5 | 31 | 58 | uni | 57 |
| 6 | fishman2x | 5 | 33 | random64-bsd | 32 | 59 | uni32 | 58 |
| 7 | gfsr4 | 6 | 34 | random64-glibc2 | 33 | 60 | vax | 59 |
| 8 | knuthran | 7 | 35 | random64-libc5 | 34 | 61 | waterman14 | 60 |
| 9 | knuthran2 | 8 | 36 | random8-bsd | 35 | 62 | zuf | 61 |
| 10 | knuthran2002 | 9 | 37 | random8-glibc2 | 36 | 63 | stdin_input_raw | 200 |
| 11 | lecuyer21 | 10 | 38 | random8-libc5 | 37 | 64 | file_input_raw | 201 |
| 12 | minstd | 11 | 39 | random-bsd | 38 | 65 | file_input | 202 |
| 13 | mrg | 12 | 40 | random-glibc2 | 39 | 66 | ca | 203 |
| 14 | mt19937 | 13 | 41 | random-libc5 | 40 | 67 | uvag | 204 |
| 15 | mt19937_1999 | 14 | 42 | randu | 41 | 68 | AES_OFB | 205 |
| 16 | mt19937_1998 | 15 | 43 | ranf | 42 | 69 | Threefish_OFB | 206 |
| 17 | r250 | 16 | 44 | ranlux | 43 | 70 | XOR (supergenerator) | 207 |
| 18 | ran0 | 17 | 45 | ranlux389 | 44 | 71 | kiss | 208 |
| 19 | ran1 | 18 | 46 | ranlxd1 | 45 | 72 | superkiss | 209 |
| 20 | ran2 | 19 | 47 | ranlxd2 | 46 | 73 | R_wichmann_hill | 400 |
| 21 | ran3 | 20 | 48 | ranlxs0 | 47 | 74 | R_marsaglia_multic. | 401 |
| 22 | rand | 21 | 49 | ranlxs1 | 48 | 75 | R_super_duper | 402 |
| 23 | rand48 | 22 | 50 | ranlxs2 | 49 | 76 | R_mersenne_twister | 403 |
| 24 | random128-bsd | 23 | 51 | ranmar | 50 | 77 | R_knuth_taocp | 404 |
| 25 | random128-glibc2 | 24 | 52 | slatec | 51 | 78 | R_knuth_taocp2 | 405 |
| 26 | random128-libc5 | 25 | 53 | taus | 52 | 79 | empty | 600 |
| 27 | random256-bsd | 26 | | | | | | |

Fig. 2. Generators available in *RDieharder*.

```
> dhtest = dieharder(rng="mt19937", test=15, psamples=100, seed=12345)
> print(dhtest)

        Diehard Runs Test

data: Created by RNG 'mt19937' with seed=12345, sample of size 100
p-value = 0.8347
```

Fig. 3. Example 1.

- The Frequency (Monobit) test: the aim of this test is to study the proportion of the number of zeros and ones in the sequence. If this were random, the ratio would be approximately the same, i.e. 1/2. The other tests on the battery should be checked once this test has been passed.

```
> dhtest = dieharder(rng=41, test=15, psamples=100000, seed=123456)
> print(dhtest)

        Diehard Runs Test

data: Created by RNG 'randu' with seed=123456, sample of size 100000
p-value <2.2e-16
```

Fig. 4. Example 2.

- Frequency test within a block: the aim of this test is to study the proportion of ones in a block of size $N$. If the sequence were random, this ratio would be expected to be $N/2$.

- The Runs test: the aim of this test is to study the runs of the sequence under study. A run of length $k$ consists of $k$ equal bits so that the pre-string bit and the $k + 1$ are different. If the sequence were random, the runs of zeros and ones would be distributed uniformly.

- Tests for the Longest-Run-of-Ones in a block: this test studies the strings of ones in a block of size $M$. It studies whether the length of the largest streak present in the sequence is consistent with what should be expected for a random sequence.

- The Binary Matrix Rank test: this test studies the ranks of the (disjoints) submatrices formed from the sequence data. If the data were random, it is expected that there would be no dependence between the sub-strings of the original sequence. This test is included in the Diehard battery.

- The Discrete Fourier Transform (Spectral) test: this test studies the peak heights in the Discrete Fourier Transform of the sequence to detect possible periodic patterns in the sequence that would imply that the sequence under study would not be random. The objective is to detect when the number of peaks exceeding the 95% threshold is statistically different from 5%. An interesting paper about this test can be seen in [31].

- The Non-overlapping Template Matching test: this test studies the number of occurrences of pre-specified target strings to see if a significant number of a non-periodic pattern occurs. In this test, a window of $m$-bits is used to study a specific pattern of size $m$. If this pattern is not found, the window is moved one position, on the contrary, if a pattern is found, the window is reset to the bit after the found pattern, and the search starts again.

- The Overlapping Template Matching test: this test studies the number of occurrences of pre-specified target strings. As in the previous test, it uses a window of $m$-bits to search for a specific $m$-bit pattern. If the pattern is not found, the window moves one bit position. Contrastingly, if a pattern is found the window moves only one bit before starting again.

- Maurer's "Universal Statistical" test: this test studies the number of bits between matching patterns. The aim of this test is to detect if one sequence can be compressed substantially without losing information. A significantly compressible sequence is considered non-random. This test was introduced by Maurer in 1992 [45] and is called universal because *"it can detect any significant deviation of a device's output statistics from the statistics of a truly random bit source when the device can be modeled as an ergodic stationary source with finite memory but arbitrary (unknown) state transition probabilities"* [45].

- The Linear Complexity test: this test studies a linear feedback shift register (LFSR). If a sequence were random it would have long LFSRs; on the contrary if the LFSR is too short, the sequence is non-random.

- The Serial test: this test studies the frequency of all possible $m$-bit patterns overlapped throughout the sequence. The aim of the test is to determine if the number of occurrences of the $2^m$ m-bit overlapping patterns is approximately the same as would be expected for a random sequence (in that case each $m$-bit pattern has the same probability of appearing). In the case $m = 1$ the frequency test is obtained.

- The Approximate Entropy test: this test studies the frequency of all possible $n$-bit patterns superimposed along the whole sequence and compares the frequencies of two consecutive blocks ($n$ and $n + 1$) with those that should appear in the case of a random sequence.

- The Cumulative Sums (Cusums) test: the random walk through the cumulative (adjusted) sum of -1 and +1 of the sequence digits, is defined. If the sequence were random, the "excursions" of the random walk should be close to 0.

- The Random Excursions test: this test studies the number of cycles that have exactly $k$ visits in a cumulative sum random walk defined as in the previous test. A cycle of a random walk consists of a sequence of steps of length one, taken at random that begin and return to the origin. The objective of the test is to determine if the number of visits to a certain state within a cycle deviates from what would be expected in the case of a random sequence.

- The Random Excursions Variant test: this test studies the total number of times a particular state is visited in a cumulative sum random walk. The aim of this test is to detect deviations from the expected number of visits to various states in the random walk.

The particular parameters that appear in the different tests of the NIST battery can be seen in table 2.

This battery has been studied in detail in [56]. In particular, its authors established the minimum lengths that the bit sequences must have in order to be used by the battery and criticized the computational times required for the verification of each of the tests.

The battery has been studied from different perspectives. For example, in [48], the authors calculate the exact non-asymptotic distribution of the $p - values$ generated by some tests included in the battery, and propose some approximations. Following the same line, [55] works for the case of test items based on the binomial distribution and discover some inconsistencies in the second level tests of the tests. They therefore propose a methodology based on the $Q - value$ as the metric of these second level tests to replace the original $p - value$ without any further modification, and the first level tests remain unchanged. Additionally, they give the correction test for the proposed second level tests based on the $Q - value$.

In order to apply this battery we should go to the *NIST* website:

$https : //csrc.nist.ðov/Projects/Random - Bit - Generation/Documentation - and - Software$ [9].

In this webpage all the information about the download and installation of this battery can be found. The source code of the files was written in ANSI C but it is possible to work under Windows operating system using an auxiliary program such as MinGW.

We will show some examples of the use of this battery. In Example 3, see Figure 5, we apply all tests of the battery of a sequence generated by generator 1 (linear congruential generator with parameters used by Fishman and Moore [20]) and 100 bitstreams.

The final result is given in a separate document .txt. In Figure 6 we show a fragment of this report. The results are represented via a table. The number of rows corresponds to the number of statistical tests applied. The number of columns, $q = 13$, are distributed as follows: columns 1-10 correspond to the frequency of $p - values$, column 11 is

GENERATOR SELECTION

[0] Input File                       [1] Linear Congruential
[2] Quadratic Congruential I         [3] Quadratic Congruential II
[4] Cubic Congruential               [5] XOR
[6] Modular Exponentiation           [7] Blum-Blum-Shub
[8] Micali-Schnorr                   [9] G Using SHA-1

Enter Choice: 1

STATISTICAL TESTS

[01] Frequency                       [02] Block Frequency
[03] Cumulative Sums                 [04] Runs
[05] Longest Runs of Ones            [06] Rank
[07] Discrete Fourier Transform      [08] Nonperiodic Template Matchings
[09] Overlapping Template Matchings  [10] Universal Statistical
[11] Approximate Entropy             [12] Random Excursions
[13] Random Excursions Variant       [14] Serial
[15] Linear Complexity

INSTRUCTIONS

Enter 0 if you DO NOT want to apply all of the
statistical tests to each sequence and 1 if you DO.

Enter Choice: 1

Parameter Adjustments

[1] Block Frequency Test - block length(M):          128
[2] NonOverlapping Template Test - block length(m):  9
[3] Overlapping Template Test - block length(m):     9
[4] Approximate Entropy Test - block length(m):      10
[5] Serial Test - block length(m):                   16
[6] Linear Complexity Test - block length(M):        500

Select Test (0 to continue): 0
How many bitstreams? 100
        Statistical Testing In Progress.............

        Statistical Testing Complete!!!!!!!!!!!!!

Fig. 5. Example 3.

the $p-value$ that arises via the application of a chi-square test, column 12 is the proportion of binary sequences that passed, and in column 13 the corresponding statistical test is shown. In this case, the sequence passes the tests.

RESULTS FOR THE UNIFORMITY OF P-VALUES AND THE PROPORTION OF PASSING SEQUENCES

generator is <Linear-Congruential>

| C1 | C2 | C3 | C4 | C5 | C6 | C7 | C8 | C9 | C10 | P-VALUE | PROPORTION | STATISTICAL TEST |
|----|----|----|----|----|----|----|----|----|-----|---------|------------|------------------|
| 6 | 15 | 8 | 9 | 11 | 7 | 10 | 11 | 13 | 10 | 0.678686 | 99/100 | Frequency |
| 11 | 7 | 11 | 6 | 10 | 13 | 15 | 13 | 6 | 8 | 0.437274 | 99/100 | BlockFrequency |
| 6 | 12 | 17 | 10 | 3 | 11 | 10 | 15 | 9 | 7 | 0.080519 | 99/100 | CumulativeSums |
| 7 | 13 | 11 | 10 | 9 | 10 | 11 | 13 | 11 | 5 | 0.779188 | 99/100 | CumulativeSums |
| 12 | 10 | 11 | 8 | 9 | 8 | 14 | 11 | 9 | 8 | 0.935716 | 97/100 | Runs |
| 8 | 8 | 10 | 18 | 7 | 9 | 13 | 10 | 11 | 6 | 0.289667 | 100/100 | LongestRun |
| 12 | 9 | 9 | 8 | 8 | 7 | 18 | 10 | 10 | 9 | 0.455937 | 99/100 | Rank |
| 5 | 11 | 12 | 11 | 13 | 5 | 15 | 13 | 8 | 7 | 0.262249 | 100/100 | FFT |
| 13 | 13 | 10 | 9 | 12 | 7 | 11 | 9 | 8 | 8 | 0.897763 | 99/100 | NonOverlappingTemplate |
| 10 | 13 | 7 | 9 | 10 | 12 | 10 | 13 | 6 | 10 | 0.851383 | 100/100 | NonOverlappingTemplate |
| 6 | 15 | 11 | 6 | 10 | 8 | 8 | 13 | 14 | 9 | 0.419021 | 100/100 | NonOverlappingTemplate |
| . | . | . | . | . | . | . | . | . | . |  |  | . |
| 6 | 7 | 4 | 3 | 5 | 6 | 7 | 7 | 5 | 6 | 0.911413 | 56/56 | RandomExcursionsVariant |
| 7 | 3 | 5 | 5 | 6 | 7 | 3 | 6 | 10 | 4 | 0.455937 | 56/56 | RandomExcursionsVariant |
| 7 | 8 | 11 | 17 | 12 | 9 | 14 | 6 | 9 | 7 | 0.275709 | 100/100 | Serial |
| 13 | 8 | 7 | 9 | 17 | 8 | 14 | 8 | 9 | 7 | 0.304126 | 99/100 | Serial |
| 9 | 12 | 6 | 10 | 15 | 13 | 7 | 11 | 9 | 8 | 0.637119 | 99/100 | LinearComplexity |

The minimum pass rate for each statistical test with the exception of the
random excursion (variant) test is approximately = 96 for a
sample size = 100 binary sequences.

The minimum pass rate for the random excursion (variant) test
is approximately = 53 for a sample size = 56 binary sequences.

For further guidelines construct a probability table using the MAPLE program
provided in the addendum section of the documentation.
- - - - - - - - - - - - - - - - - - - - - - - - - - - - - - - - - - - - - - -

Fig. 6.  Example 3. Fragment of the final analysis report for the sequence generated by the *LCG*.

Other example in which the sequence passes the test is Example 4, see Figure 7. In this case we apply the NIST
battery to an external file uploaded by the user. We use a sequence of numbers generated by *dev/urandom* and 100
bitstreams. In order to do this we need to choose option 0 in the initial menu of the program.

A fragment of the final analysis report of Example 4 can be seen in Figure 8.

Now we will use another sequence generated by a biased generator. We generate a sequence by using *urandom* in
Python, and then we delete some occurrences of a certain number, so by its construction, this sequence is not random.
We will apply NIST battery and we will see the results in Example 5, see Figure 9.

A fragment of the final analysis report can be seen in Figure 10. Results which NIST interprets as non-randomness of
the data are marked by an asterisk.

C: \Users\AL\Documents\ sts-2_1_2 \ sts-2.1.2 > assess.exe 1000000

<u>GENERATOR SELECTION</u>

        [0] Input File            [1] Linear Congruential
        [2] Quadratic Congruential I    [3] Quadratic Congruential II
        [4] Cubic Congruential       [5] XOR
        [6] Modular Exponentiation   [7] Blum-Blum-Shub
        [8] Micali-Schnorr        [9] G Using SHA-1

Enter Choice: 0

    User Prescribed Input File: C: \Users\AL\n0

<u>STATISTICAL TESTS</u>

        [01] Frequency         [02] Block Frequency
        [03] Cumulative Sums     [04] Runs
        [05] Longest Runs of Ones    [06] Rank
        [07] Discrete Fourier Transform  [08] Nonperiodic Template Matchings
        [09] Overlapping Template Matchings [10] Universal Statistical
        [11] Approximate Entropy    [12] Random Excursions
        [13] Random Excursions Variant  [14] Serial
        [15] Linear Complexity

  INSTRUCTIONS

      Enter 0 if you DO NOT want to apply all of the
      statistical tests to each sequence and 1 if you DO.

Enter Choice: 1

        [1] Block Frequency Test - block length(M):     128
        [2] NonOverlapping Template Test - block length(m):  9
        [3] Overlapping Template Test - block length(m):   9
        [4] Approximate Entropy Test - block length(m):   10
        [5] Serial Test - block length(m):          16
        [6] Linear Complexity Test - block length(M):    500

Select Test (0 to continue): 0
How many bitstreams? 100

Input File Format:
        [0] ASCII-A sequence of ASCII 0's and 1's
        [1] Binary-Each byte in data file contains 8 bits of data

Select input mode: 1

    Statistical Testing In Progress............

    Statistical Testing Complete!!!!!!!!!!!!

Fig. 7. Example 4.

RESULTS FOR THE UNIFORMITY OF P-VALUES AND THE PROPORTION OF PASSING SEQUENCES

generator is <Linear-Congruential>

| C1 | C2 | C3 | C4 | C5 | C6 | C7 | C8 | C9 | C10 | P-VALUE | PROPORTION | STATISTICAL TEST |
|----|----|----|----|----|----|----|----|----|-----|---------|------------|------------------|
| 9  | 11 | 13 | 11 | 5  | 12 | 7  | 12 | 15 | 5   | 0.319084 | 97/100  | Frequency |
| 9  | 13 | 6  | 8  | 11 | 8  | 14 | 6  | 14 | 11  | 0.494392 | 98/100  | BlockFrequency |
| 11 | 11 | 9  | 14 | 13 | 8  | 13 | 3  | 6  | 12  | 0.275709 | 97/100  | CumulativeSums |
| 10 | 12 | 12 | 10 | 6  | 10 | 9  | 9  | 13 | 9   | 0.935716 | 98/100  | CumulativeSums |
| 6  | 8  | 16 | 7  | 11 | 13 | 6  | 9  | 12 | 12  | 0.350485 | 100/100 | Runs |
| 8  | 4  | 12 | 10 | 11 | 9  | 12 | 14 | 10 | 10  | 0.678686 | 99/100  | LongestRun |
| 12 | 12 | 12 | 9  | 10 | 11 | 12 | 9  | 5  | 8   | 0.851383 | 98/100  | Rank |
| 9  | 12 | 12 | 11 | 11 | 7  | 11 | 9  | 13 | 5   | 0.779188 | 99/100  | FFT |
| 13 | 11 | 10 | 12 | 14 | 9  | 9  | 6  | 10 | 6   | 0.699313 | 98/100  | NonOverlappingTemplate |
| 11 | 10 | 5  | 11 | 10 | 8  | 10 | 11 | 16 | 8   | 0.616305 | 99/100  | NonOverlappingTemplate |
| 10 | 11 | 3  | 15 | 9  | 9  | 9  | 8  | 13 | 13  | 0.350485 | 98/100  | NonOverlappingTemplate |
| .  | .  | .  | .  | .  | .  | .  | .  | .  | .   | .        |            | . |
| 8  | 6  | 3  | 4  | 5  | 3  | 5  | 3  | 8  | 9   | 0.383827 | 53/54   | RandomExcursionsVariant |
| 7  | 7  | 1  | 2  | 9  | 2  | 8  | 5  | 6  | 7   | 0.108791 | 53/54   | RandomExcursionsVariant |
| 7  | 3  | 5  | 4  | 3  | 6  | 3  | 5  | 11 | 7   | 0.236810 | 52/54   | RandomExcursionsVariant |
| 5  | 5  | 6  | 3  | 1  | 9  | 4  | 7  | 8  | 6   | 0.319084 | 52/54   | RandomExcursionsVariant |
| 4  | 7  | 2  | 7  | 4  | 5  | 5  | 6  | 8  | 6   | 0.739918 | 52/54   | RandomExcursionsVariant |
| 10 | 9  | 15 | 8  | 7  | 6  | 15 | 12 | 8  | 10  | 0.455937 | 100/100 | Serial |
| 9  | 8  | 8  | 4  | 13 | 20 | 8  | 9  | 11 | 10  | 0.066882 | 99/100  | Serial |
| 13 | 10 | 9  | 4  | 11 | 10 | 12 | 9  | 9  | 13  | 0.719747 | 100/100 | LinearComplexity |

The minimum pass rate for each statistical test with the exception of the
random excursion (variant) test is approximately = 96 for a
sample size = 100 binary sequences.

The minimum pass rate for the random excursion (variant) test
is approximately = 51 for a sample size = 54 binary sequences.

For further guidelines construct a probability table using the MAPLE program
provided in the addendum section of the documentation.

Fig. 8.  Example 4. Fragment of the final analysis report for the sequence n0 generated by *dev/urandom*.

## 3.5   TestU01 battery

L'Ecuyer and Simard [36] proposed a battery called TestU01 implemented in C that is easier to apply from a computational point of view and that includes many of the main tests in the literature. It is a very extensive battery that includes many of the Diehard and NIST tests and other tests that discover problems in some generators that pass Diehard and NIST batteries successfully. It is a more flexible battery than the previous ones, the implementations are more efficient, and it

C: \Users\AL\Documents\ sts-2_1_2 \ sts-2.1.2 > 8000000

GENERATOR SELECTION

[0] Input File                          [1] Linear Congruential
[2] Quadratic Congruential I            [3] Quadratic Congruential II
[4] Cubic Congruential                  [5] XOR
[6] Modular Exponentiation              [7] Blum-Blum-Shub
[8] Micali-Schnorr                      [9] G Using SHA-1

Enter Choice: 0

User Prescribed Input File: C: \Users\AL\epsilon0_0.bin

STATISTICAL TESTS

[01] Frequency                          [02] Block Frequency
[03] Cumulative Sums                    [04] Runs
[05] Longest Runs of Ones               [06] Rank
[07] Discrete Fourier Transform         [08] Nonperiodic Template Matchings
[09] Overlapping Template Matchings     [10] Universal Statistical
[11] Approximate Entropy                [12] Random Excursions
[13] Random Excursions Variant          [14] Serial
[15] Linear Complexity

INSTRUCTIONS

Enter 0 if you DO NOT want to apply all of the
statistical tests to each sequence and 1 if you DO.

Enter Choice: 1

Parameter Adjustments

[1] Block Frequency Test - block length(M):              128
[2] NonOverlapping Template Test - block length(m):        9
[3] Overlapping Template Test - block length(m):           9
[4] Approximate Entropy Test - block length(m):          10
[5] Serial Test - block length(m):                       16
[6] Linear Complexity Test - block length(M):           500

Select Test (0 to continue): 0
How many bitstreams? 100

Input File Format:
[0] ASCII-A sequence of ASCII 0's and 1's
[1] Binary-Each byte in data file contains 8 bits of data

Select input mode: 1

Statistical Testing In Progress...........

Statistical Testing Complete!!!!!!!!!!!!

Fig. 9.  Example 5.

RESULTS FOR THE UNIFORMITY OF P-VALUES AND THE PROPORTION OF PASSING SEQUENCES

generator is <Linear-Congruential>

| C1 | C2 | C3 | C4 | C5 | C6 | C7 | C8 | C9 | C10 | P-VALUE | PROPORTION | STATISTICAL TEST |
|----|----|----|----|----|----|----|----|----|-----|---------|------------|------------------|
| 1  | 92 | 1  | 0  | 0  | 0  | 2  | 1  | 1  | 2   | 0.000000 * | 99/100  | Frequency |
| 0  | 2  | 2  | 0  | 91 | 0  | 2  | 0  | 2  | 1   | 0.000000 * | 100/100 | BlockFrequency |
| 1  | 91 | 2  | 1  | 1  | 0  | 3  | 0  | 0  | 1   | 0.000000 * | 99/100  | CumulativeSums |
| 1  | 91 | 1  | 1  | 2  | 0  | 0  | 1  | 3  | 0   | 0.000000 * | 99/100  | CumulativeSums |
| 1  | 0  | 92 | 0  | 1  | 2  | 1  | 1  | 0  | 2   | 0.000000 * | 99/100  | Runs |
| 91 | 2  | 2  | 1  | 1  | 0  | 2  | 0  | 1  | 0   | 0.000000 * | 100/100 | LongestRun |
| 0  | 0  | 1  | 91 | 0  | 1  | 1  | 4  | 0  | 2   | 0.000000 * | 100/100 | Rank |
| 0  | 1  | 92 | 1  | 2  | 1  | 1  | 1  | 1  | 0   | 0.000000 * | 100/100 | FFT |
| 1  | 2  | 91 | 0  | 1  | 0  | 2  | 0  | 2  | 1   | 0.000000 * | 99/100  | NonOverlappingTemplate |
| 1  | 2  | 90 | 1  | 2  | 2  | 1  | 1  | 0  | 0   | 0.000000 * | 100/100 | NonOverlappingTemplate |

⋮    ⋮    ⋮       ⋮    ⋮    ⋮    ⋮    ⋮    ⋮        ⋮              ⋮

The minimum pass rate for each statistical test with the exception of the
random excursion (variant) test is approximately = 96 for a
sample size = 100 binary sequences.

The minimum pass rate for the random excursion (variant) test
is approximately = 7 for a sample size = 8 binary sequences.

For further guidelines construct a probability table using the MAPLE program
provided in the addendum section of the documentation.

Fig. 10. Example 5. Fragment of the final analysis report for the biased sequence.

can support larger sample sizes and a wider range of test parameters. Tests are available for bit strings and for real number sequences in the interval $(0, 1)$.

It consists of six test batteries, three of which are oriented to sequences of values within the interval $(0, 1)$ and are used to test the structure and output of the PRNGs (Small Crush, Crush, Big Crush) and the other three oriented for analyzing bit sequences (Rabbit, Alphabit and BlockAlphabit-the latter applies the Alphabit test battery repeatedly to a generator or a binary file after rearranging the bits by blocks of different sizes (2, 4, 8, 16, 32 bits)-). According to Gentle [23], the time needed to test a single generator with Big Crush can be more than ten hours on a high-end PC. However, with Small Crush it is two orders of magnitude faster, so it is advisable to start testing with Small Crush and to stop if the generator does not pass the tests. Otherwise, it would continue with Crush. If the generator does not pass the tests it will stop, otherwise, it will continue with Big Crush. On a 64-bit AMD Athlon processor running at 2.4 GHz, Rabbit takes about 2 seconds to test a 220-bit stream in a file and about 9 minutes for a 230-bit stream. Alphabit takes less than 1 second for 220-bit and about 1 minute for 230-bit [36]. The tests included in each suite in TestU01 battery can be seen in Table 4.

This battery can be downloaded in:

*http://simul.iro.umontreal.ca/testu01/tu01.html* [37].

In this webpage we can find the source files, the installation and user's guide. This battery is implemented in the ANSI C language but as in the previous subsection, it can be used in Windows. In case you want to work under Windows operating system you have to be careful to export the libraries correctly and this depends on your operating system version.

Now we will show some examples of the use of this battery. First of all we show Example 6. In this case we use an internal example of the package TestU01, *birth1.c*. In this example we use a *LCG* with parameters $m$ = 2147483647, $a$ = 397204094, $b$ = 0 and seed $s$ = 12345, the result we obtain of the analysis can be seen in Figure 11. It is important to see how the situation changes if the number of births does if. If $n$ = 1000, the $p - value$ is 0.04 > $\alpha$ and when $n$ = 10000, $p - value < \alpha$.

Now we modify the file *birth1.c* and we include the library *bbattery.h* in order to apply Small Crush. **This** is our Example 7. The results can be seen in Figure 12.

In Example 8 we apply Small Crush to a new file. In this case we use a new generator by combining two *LCGs* (using the function *unif01_CreateCombAdd2*), the first one with parameters $m$ = $2^{31} - 1$, $a$ = 630360016, $b$ = 0 and seed $s$ = 12345 (used in some FORTRAN) and the other with parameters $m$ = 2147483563, $a$ = 16807, $b$ = 0, and $s$ = 12345 (used for example in APL, IMSL and SIMPL/I). The results are shown in Figure 13, the program prints a convenient summary: the names of all tests that produce a $p - value$ outside the [0.1, 0.99] interval.

### 3.6 ENT battery

There are other less used batteries such as ENT battery, proposed by Walker [65], which is used by the random.org website. ENT claims to test for simulation and cryptographic purposes. ENT battery has two modes of working binary and byte, depending on the mode, different statistics will be calculated and displayed. This battery includes the following tests (Observation: from the statistical point of view, there is one test, $\chi^2$ and the others are measures indicators):

- Entropy: this test calculates the entropy of the sequence under study. If the sequence were random, it should be rich in entropy. The maximum theoretical entropy of a long sequence is 8 bits per byte, in byte mode, and one bit per bit in binary mode.
- Chi-square Test: this test computes the frequency of the symbols and compares it with the frequency expected in a uniform distribution this is done by a $\chi^2$ statistic.
- Arithmetic Mean: the arithmetic mean of the sequence values is calculated. If the sequence were random, the expected value would be in 0.5 in binary mode and 127.5 in byte mode.
- Monte Carlo Value for $\pi$: in this test, the sequence of numbers is interpreted as coordinates in a square and the number of points that fall within the circle inscribed in that square is counted. That number is taken as an estimation of $\pi$. If the sequence were not random, the approximation to the true value would not be good.
- Serial Correlation Coefficient: the correlation between two consecutive symbols (bits or bytes) is calculated. This correlation is expected to be low in random sequences.

This battery can be downloaded in:

*https://www.fourmilab.ch/random/* [65].

Installation instructions (to be done on Unix and Windows) can be found and the guidelines for its use and interpretation.

```
$ birth1.exe
''''''''''''''''''''''''''''''''''''''''''''''''''''''''''''''''''

HOST =

ulcg_CreateLCG: m = 2147483647, a = 397204094, c = 0, s = 12345

smarsa_BirthdaySpacings test:
__________________________________________
   N = 1, n = 1000, r = 0, d = 10000, t = 2, p = 1

      Number of cells = d^t = 100000000
      Lambda = Poisson mean = 2.5000
__________________________________________
         Total expected number = N*Lambda: 2.50

         Total observed number: 6

         p-value of test: 0.04
__________________________________________
CPU time used : 00:00:00.00

Generator state:
s = 1858647048

''''''''''''''''''''''''''''''''''''''''''''''''''''''''''''''''''
HOST =

ulcg_CreateLCG: m = 2147483647, a = 397204094, c = 0, s = 12345

smarsa_BirthdaySpacings test:
__________________________________________
   N = 1, n = 10000, r = 0, d = 1000000, t = 2, p = 1

      Number of cells = d^t = 1000000000000
      Lambda = Poisson mean = 0.2500
__________________________________________
         Total expected number = N*Lambda: 0.25

         Total observed number: 44

         p-value of test: 9.5e-82 *****
__________________________________________
CPU time used : 00:00:00.00

Generator state:
s = 731506484
```

Fig. 11. Example 6. Results of the analysis of *birth1.c* file.

```
========= Summary results of SmallCrush =========

Version: TestU01 1.2.3
Generator: ulcg_CreateLCG
Number of statistics: 15
Total CPU time: 00:00:25.23
The following tests gave p-values outside [0.001,0.9990]:
(eps means a value <1.0e-300):
(eps1 means a value <1.0e-015):
  Test                      p-value
  1 BirthdaySpacings         eps
  2 Collision               1-eps1
All other test were passed
```

Fig. 12. Example 7. Results of the analysis of the modified *birth1.c* file.

```
========= Summary results of SmallCrush =========

Version: TestU01 1.2.3
Generator: My Combination of LCGs, unif01_CreateComAdd2
Number of statistics: 15
Total CPU time: 00:00:30.39
The following tests gave p-values outside [0.001,0.9990]:
(eps means a value <1.0e-300):
(eps1 means a value <1.0e-015):
  Test                      p-value
  1 BirthdaySpacings   1.6e-118
All other test were passed
```

Fig. 13. Example 8. Results of the analysis of the combining generator.

Now we show Example 9. This is an application of the ENT battery to an internal file available in its designed software, the sequence *ent.c*. This file has many non-random patterns (for example, the character space will be extremely overrepresented, and reserved words like int will be repeated many times). The results are shown in Figure 14.

Now we apply the ENT battery for a sequence of 1000000 numbers generated by using a *LCG* with parameters $m = 2^{48}$, $a = 25214903917$ and $b = 11$ with seed 123456789, Example 10. The results of the application of ENT battery can be seen in Figure 15 and as it can be seen the sequence is not good from the perspective of randomness.

In Example 11 we use the sequence n0 previously used in Example 4 and we apply ENT battery. The results can be seen in Figure 16. As it can be seen these results are compatible with the randomness of the sequence.

Despite the speed and simplicity of the ENT battery, it has certain weaknesses. For instance, it presents some problems of dependence between tests (for example Entropy vs Chi-square Test among others) and problems related to poorly designed generators with the presence of serial correlation in which the battery does not detect that the numbers obtained are "bad" (degenerated by design). Let's show this in Example 12. In this case we use the generator

Entropy = 4.846849 bits per byte.
Optimum compression would reduce the size of this 7989 byte file by 39 percent.
Chi square distribution for 7989 samples is 158914.00, and randomly would exceed this value less than 0.01 percent of the times.
Arithmetic mean value of data bytes is 73.3497 (127.5 = random).
Monte Carlo value for Pi is 4.000000000 (error 27.32 percent).
Serial correlation coefficient is 0.471176 (totally uncorrelated = 0.0).

Fig. 14. Example 9. ENT results for the sequence *ent.c*.

Entropy = 3.653481 bits per byte.
Optimum compression would reduce the size of this 16000002 byte file by 54 percent.
Chi square distribution for 16000002 samples is 356045004.10, and randomly would exceed this value less than 0.01 percent of the times.
Arithmetic mean value of data bytes is 50.3503 (127.5 = random).
Monte Carlo value for Pi is 4.000000000 (error 27.32 percent).
Serial correlation coefficient is 0.135956 (totally uncorrelated = 0.0).

Fig. 15. Example 10. ENT results for the sequence *u* generated by a *LCG*.

Entropy = 7.999987 bits per byte.
Optimum compression would reduce the size of this 15000000 byte file by 0 percent.
Chi square distribution for 15000000 samples is 263.59, and randomly would exceed this value less than 34.26 percent of the times.
Arithmetic mean value of data bytes is 127.5251 (127.5 = random).
Monte Carlo value for Pi is 3.140208000 (error 0.04 percent).
Serial correlation coefficient is -0.000196 (totally uncorrelated = 0.0).

Fig. 16. Example 11. ENT results for the sequence n0 generated by *dev/urandom*.

defined in Algorithm 1 (which generates non-random sequences) we can pass with good results the ENT battery as we can see in Figure 17.

Entropy = 7.999912 bits per byte.
Optimum compression would reduce the size of this 2097152 byte file by 0 percent.
Chi square distribution for 2097152 samples is 256.00, and randomly would exceed this value less than 47.06 percent of the times.
Arithmetic mean value of data bytes is 127.5005 (127.5 = random).
Monte Carlo value for Pi is 3.140925540 (error 0.02 percent).
Serial correlation coefficient is 0.000000 (totally uncorrelated = 0.0).

Fig. 17. Example 12. ENT results for a biased sequence generated by Algorithm 1.

```
for 1 ≤ v ≤ 16 do
    for 0 ≤ i ≤ 255 do
        for 0 ≤ j ≤ 255 do
            if j = 250 and i  mod 4 = 0 then
                output ← i;
                output ← 251;
                else output ← i;
                    output ← j;
            end
        end
    end
```

**Algorithm 1:** Degenerated generator

Now it is an issue of interest to show a comparison of the described batteries. In Table 5 we summarize some advantages and disadvantages of the previous test batteries.

### 3.7 Other batteries of statistical tests

Other batteries are for example the one proposed by Vattulainen et al. [63], Cryp-X [7] or SPRNG [44]. Vattulainen et al. in 1995 proposed a battery based on the ISING model and random walks on lattices, including the Autocorrelation, Cluster, n-block and random walk tests. The Cryp-X battery was introduced by the Information Security Research Center at Queensland University of Technology for commercial purposes. This battery includes 6 tests: Binary derivative, Change point, Frequency, Linear complexity, Runs and Sequence complexity. Crypt-X supports stream ciphers, block ciphers and keystream generators. The Scalable Parallel Random Number Generators Library (SPRNG) battery implements a number of "parallel" versions of the tests proposed in the Knuth battery [34], as well as the "inherently parallel" sum-of-independent-distributions test. It also implements other tests, based on the ISING model and random walk tests [43]. The SPRNG battery is suitable for applications in parallel Monte Carlo simulation. This battery is widely used, some institutions that use it are (see *http://www.sprng.org/*): Aeronautical Systems Center Major Shared Resource Center, Department of Defense, Los Alamos National Lab, National Center for Computational Sciences, Computer Services for Academic Research (CSAR) of UK National High Performance Computing (HPC) Service, United Kingdom, Irish Centre for High-End Computing, Colorado School of Mines, University of California - San Diego, Ateneo de Manila University, Philippines among others.

A comparative between tests included in the previous described batteries can be seen in Table 3.

In addition to the previously mentioned tests, there are others which are not included in the explained batteries. For example, the SAC (Strict Avalanche Criterion) test proposed by Hernández et al. [29] the adaptive $\chi^2$ test the Book Stack and Order tests propose by Ryabko et. al. ([54], [53]), randomness test based on random walk proposed by Doganaksoy et. al. [12] or based on Golomb's postulates [14] and topological binary test proposed by [2]. An interesting paper that uses and describes some of these test is [11].

## 4 SOME RECOMMENDATIONS

As described above, there are numerous test batteries to be applied, being the most recognized the Dieharder, the NIST and the TestU01. Many authors have studied different generators, properties, etc., by using these principal batteries, see for example [38], [4] or [42], in which the authors use these batteries to verify the goodness of the (random) generated sequences (in [42], the authors also use ENT and Diehard batteries). Sometimes, there are only a few tests of a battery which are applied, for example, in [5] the authors apply the NIST tests: Frequency test, Cumulative Sums test and Runs test. In [30] the authors use TestU01 battery for testing randomness of the pseudo-random numbers generated by chaotic maps, and, in order to obtain more neutral test results, in [61] the authors use NIST battery in order to verify the goodness of the sequences they obtain by using a QRNG.

In other cases there is some particular test or measure of these batteries in which authors are interested. For example, in [66] the authors review and discuss the entropy estimators including the NIST SP 800-90B estimators and neural network based estimators paying attention to their limitations on the time-varying data. In addition, they propose a scheme for the estimation of the entropy adopting change detection strategies to deal with this problem.

Which tests are the good ones? Which is the best battery? Do we have to use the same tests for true random numbers and for pseudo-random numbers? These questions have no general answer. The essential point is to be able to find a criterion that will tell us which battery is better, when to use one instead of the other, if there are substantial differences between them, etc. Some of these questions have been dealt with in previous lines. We have described the tests that are included in each of them, and we have been able to verify that many of them are present in several of them, and we have indicated which test should be passed as a necessary (but not sufficient) condition of randomness (runs test) or in what order, for example, one should work with TestU01. Unfortunately, no defined criteria exist to be able to answer the other questions raised, such as which battery is better. From a practical perspective and given that to obtain quality certification of products it is necessary the qualification of the US National Institute of Standards and Technology, we consider the use of the battery they propose of great importance (although, as we will see later, it also has some aspects that need to be improved). On the other hand, the Dieharder battery in its latest version has incorporated numerous tests included in the NIST battery and intends in future updates to include the entire NIST battery in addition to any tests considered relevant for testing randomness. From our point of view, the objective is good, but it does not go into detail about whether all the tests are actually necessary, whether there may be redundancies between them, etc. In other words, it is not enough to build a compendium of all the tests that can be defined, but rather an analysis of them should be carried out in order to create a battery that contains the right and necessary number of tests that should be used. TestU01 also implements some tests included in other batteries such as Diehard or NIST batteries, and it can be considered a kind of platform on which almost all test could be implemented. Also, TestU01 battery is more flexible, the mode of implementation is more efficient, the battery can work with samples of big sizes and allows the user to work with more test parameters than other batteries.

That is why we will focus on the NIST and TestU01 batteries, but with some observations to take into account.

NIST provides guidance ([52], section 5) on how to interpret the results obtained after the application of the tests and states that in order for the data to be considered random, all tests on the battery must be passed. This condition is too rigid since, among other situations, truly random numbers have a high probability (80%) of failing at least one of the NIST battery tests. On the other hand, there are some statistical reasons that allow us to relax this restriction we will explain later.

NIST battery process the input data and gives the results of the analysis in several documents. We will pay attention to the Final Analysis Report file. This is a document .txt in which it appears the following information ([52], section 5.7.):

- Ten columns whose entries correspond to the number of *p-values* that fall within intervals: $[0, 0.1), [0.1, 0.2),...,[0.9, 1)$.

- One column named P-value in which the result for uniformity testing *p-values* computed for a given test are presented.
- One column named Proportion in which the proportion of sequences that pass a given test is presented.
- One column named Test in which the applied test is noticed.

The *p-values* computed by a single test should be distributed as a $U(0,1)$ random variable. NIST uses one sample test $\chi^2$ to verify the uniformity of the *p-values*. Note that $\chi^2$ test works well when the number of tested sequences is at least 55. The values in the P-value column represent the *p-value* of this uniformity test. A computed small *p-value* indicates a problem of the generator, but it is very difficult to identify exactly which problem it is. The NIST battery documentation recommends taking a level $\alpha = 0.0001$ for this test. As it is known, the smaller the value of $\alpha$, the higher the value of $\beta$ (Type-II error). From a practical point of view, a small value of $\beta$ is more important than a small value of $\alpha$, that is, the probability of accepting a bad generator is wanted to be small (we need small values of $\beta$). Hence, from this perspective, we consider that NIST requirement should be relaxed and consider values $\alpha = 0.01$ or $\alpha = 0.001$.

In column Proportion, the proportion of sequences that pass a test is represented. The probability that a random sample passes a test is equal to $1 - \alpha$, for multiple sequences this probability can be approximated by $1 - \alpha$. According to the NIST battery documentation ([52], section 4.2.1.), the proportion of passing sequences should fall into the interval:

$$\left( (1-\alpha) - 3\sqrt{\frac{\alpha(1-\alpha)}{k}}, (1-\alpha) + 3\sqrt{\frac{\alpha(1-\alpha)}{k}} \right) \tag{1}$$

being $k$ the number of tested sequences. This interval is derived by using the Moivre-Laplace theorem which assesses that the statistic $p$ (the approximation of the proportion) has an asymptotic $N(\mu, \hat{\sigma})$ distribution, where $\hat{\sigma} = \sqrt{\frac{p(1-p)}{k}}$ is the (estimated) standard deviation. In this case, if we set a confidence level $1 - \alpha$, a constant $l$ such that $P[-l \leq Z \leq l] = 1 - \alpha$ with $Z \sim N(0,1)$, can be found in order to define the interval:

$$\left( p - l\sqrt{\frac{p(1-p)}{k}}, p + l\sqrt{\frac{p(1-p)}{k}} \right) \tag{2}$$

Hence, in the NIST proposal $l = 3$ and its associate $\alpha$ level is 0.0028 (Type-I error), and note that $0.0028 < 0.01$ that is the interval that NIST gives does not reflect a confidence of 99%. If we want t work with a value $\alpha = 0.01$ the correct selection of $l$ should be 2.5758 and so the interval is:

$$\left( (1-\alpha) - 2.5758\sqrt{\frac{\alpha(1-\alpha)}{k}}, (1-\alpha) + 2.5758\sqrt{\frac{\alpha(1-\alpha)}{k}} \right) \tag{3}$$

The interpretation of the result is that in the 99% of the cases the proportion should fall into the confidence interval (3) and, so, the critic region of the hypothesis test would be the complementary of that interval with probability $\alpha = 0.01$. In addition, we consider that it is important to pay attention to some problems in the use of the discrete Fourier transform test. Following this point, in [47] the authors study in detail this test and point out that the most crucial problem is that its reference distribution from the statistical test is not derived mathematically, which is estimated numerically and it is concluded that this test should not be used unless the reference distribution is derived mathematically. Additionally, the authors propose a test, whose reference distribution is mathematically derived.

With these observations, we believe that the use of the NIST battery with this interpretation of the results would be more appropriate and statistically adjusted.

On the other hand, taking up the subject of possible dependence between tests, it is worth highlighting some studies whose main objective is precisely to detect them, for example in [15] or [13]. At present, this is a line of research that is being worked on.

TestU01, on the other hand, provides a very powerful platform where the user can select generators, specific statistical tests, complete pre-defined batteries and tools to apply tests to particular generators. This is an attempt to cover the bibliography and methodologies that have been designed over time. As mentioned before, the problem of execution and obtaining results is present and that is why it is recommended following a specific order when executing it.

The correspondence between tests in NIST battery and TestU01 battery can be seen in Table 6 where NIST battery tests and their corresponding functions on TestU01 battery are shown, in case there is no exact relation, similar functions are included, the latter is indicated with (*).

## 5  CONCLUSIONS

In this paper different hypothesis test batteries existing in the literature have been described in detail oriented to the verification of the desirable statistical properties for a data sequence either obtained with a RNG or a PRNG.

The batteries are composed by different hypothesis tests that can be applied separately or together with the rest of the battery. On the other hand, among the existing batteries it can be checked how in many cases the same tests are applied. In others the number of tests to be used is extended or even some tests are changed for others generating new batteries. When applying a certain battery, the initial problems arise in selecting the level of significance of the tests that is essential for the rejection or not of the null hypothesis under study. Generally, this level is set at $\alpha = 0.01$ or $\alpha = 0.05$, and values lower than 0.01 make it very difficult to reject the null hypothesis and, therefore, the probability of deciding incorrectly on the randomness of the sequence increases. An interesting aspect would be to study how the response of the decision maker would vary, that is, how the rejection or not of the null hypothesis would vary depending on the variation of the $\alpha$ significance level and its statistical implications.

On the other hand, as seen throughout this paper, in many of the batteries, the tests are designed under certain parameters that in some cases are preset, avoiding their modification by the analyst. This last one causes certain rigidity in its application and important statistical consequences as for example to increases in Type-I error. In this paper the different pros and cons of each of the existing batteries have been discussed. Moreover, we have given some recommendations abut the use of the batteries.

Two main lines of research derived from our review can be extracted: the first is based on carrying out studies on the dependence or not of the different tests included in the batteries. The second and as a consequence of the above, is the design of efficient test batteries in which there is no overlap in the checks of the statistical properties.

### ACKNOWLEDGEMENTS

This work has received funding from THEIA (Techniques for Integrity and Authentication of Multimedia Files of Mobile Devices) UCM project (FEI-EU-19-04).

| NIST TESTS | Recommended sample size n (number of bites) |
|---|---|
| The Frequency (Monobit) test | $n \geq 100$ |
| Frequency test within a block | $n \geq 100$ <br> $n \geq MN$ <br> $M \geq 20,\ M \geq 0.01n$ <br> $N < 100$ <br> Being $M$ the block size and $N = int\,(n/M)$ |
| The Runs test | $n \geq 100$ |
| Tests for the Longest-Run-of-Ones in a block | If $n = 128$ then $M = 8$ <br> If $n = 6272$ then $M = 128$ <br> If $n = 750000$ then $M = 10^4$ <br> Being $M$ the block size pre-set in the test to accommodate to values 8, 128 and $10^4$ |
| The Binary Matrix Rank test | $n \geq 38MQ$ <br> Being $M$ the number of rows of the matrix and $Q$ the number of columns <br> It is consider values $M = Q = 32$, so each sequence to be tested should consist of a minimum of 38912 bits |
| The Discrete Fourier Transform (Spectral) test | $n \geq 1000$ |
| The Non-overlapping Template Matching test | $m = 2 \ldots 10$ preferably $m = 9$ or $m = 10$ <br> Being $m$ the length in bits of each template <br> $M > 0.01n$ being $M$ the length in bits of the substring to be tested <br> $N = int\,(n/M)$ <br> $N \leq 100$ <br> Being $N$ the number of independent blocks, initially fixed at $N = 8$ |
| The Overlapping Template Matching test | $n \geq 10^6$ <br> $m = 9$ or $m = 10$ being the length in bits of the template <br> $n \geq NM$, being $M$ the length in bits of a substring of to be tested <br> $M = 1032$ in the test code <br> Being $N$ the number of independent blocks of $n$ <br> $N = 968$ in the test code |
| Maurer's "Universal Statistical" test | The sequence of length $n$ is divided into two segments of $L$-bit blocks <br> The first segment consists of $Q$ initialization blocks and the second one consists of $K$ test blocks <br> $n \geq (Q + K)L$ with $6 \leq L \leq 16$, $Q = 10 * 2^L$, $K \approx 1000 \cdot 2^L$ <br> In particular the following recommendations are given: <br><br> <table><tr><td>$n$</td><td>$L$</td><td>$Q$</td></tr><tr><td>$\geq 387840$</td><td>6</td><td>640</td></tr><tr><td>$\geq 904960$</td><td>7</td><td>1280</td></tr><tr><td>$\geq 2068480$</td><td>8</td><td>2560</td></tr><tr><td>$\geq 4654080$</td><td>9</td><td>5120</td></tr><tr><td>$\geq 10342400$</td><td>10</td><td>10240</td></tr><tr><td>$\geq 22753280$</td><td>11</td><td>20480</td></tr><tr><td>$\geq 49643520$</td><td>12</td><td>40960</td></tr><tr><td>$\geq 107560960$</td><td>13</td><td>81920</td></tr><tr><td>$\geq 231669760$</td><td>14</td><td>163840</td></tr><tr><td>$\geq 496435200$</td><td>15</td><td>327680</td></tr><tr><td>$\geq 1059061760$</td><td>16</td><td>655360</td></tr></table> |
| The Linear Complexity test | $n \geq 106$ <br> $500 \leq M \leq 5000$ <br> $N \geq 200$ <br> Being $M$ the length in bits of a block <br> Being $N$ the number of independent blocks of $M$ bits, $n = MN$ |
| The Serial test | $m < int\,(\log_2 n) - 2$ <br> Being $m$ the length in bits of each block |
| The Approximate Entropy test | $m < int\,(\log_2 n) - 5$ <br> Being $m$ the length of each block, in this case, the first block length used in the test. m+1 is the second block length used |
| The Cumulative Sums (Cusums) test | $n \geq 100$ |
| The Random Excursions test | $n \geq 10^6$ |
| The Random Excursions Variant test | $n \geq 10^6$ |

Table 2. NIST tests and parameters values.

| Test \ Battery | Knuth | Diehard | Dieharder | NIST | ENT | Crypt-X | SPRNG |
|---|---|---|---|---|---|---|---|
| Frequency | X | | X | X | | X | |
| Serial | X | | | X | | | X |
| Gap | X | | | | | | X |
| Poker | X | | | | | | X |
| Coupon collector's | X | | | | | | X |
| Permutation | X | | | | | | X |
| Runs | X | X | X | X | | X | X |
| Maximum of t | X | | | | | | X |
| Collision | X | | | | | | X |
| Birthday Spacings | X | X | X | | | | |
| Serial Correlation | X | | | | X | | |
| Subsequences | X | | | | | | |
| OPERM5 | | X | X | | | | |
| Binary rank | | X | X | X | | | |
| Bitstream | | X | X | | | | |
| OPSO, OQSO and DNA | | X | X | | | | |
| The count-the-1's on a stream of bytes | | X | X | | | | |
| The count-the-1's for specific bytes | | X | X | | | | |
| The parking lot | | X | X | | | | |
| Minimum distance | | X | X | | | | |
| 3D spheres | | X | X | | | | |
| The squeeze | | X | X | | | | |
| The overlapping sums | | X | X | | | | |
| The craps | | X | X | | | | |
| K-S | | | X | | | | |
| Greatest Common Divisor Marsaglia-Tsang | | | X | | | | |
| Generalized minimum distance | | | X | | | | |
| Lagged sum | | | X | | | | |
| Permutations | | | X | | | | |
| Generalized Serial | | | X | | | | |
| Bit distribution | | | X | | | | |
| Frequency within a block | | | | X | | | |
| Longest-Run-of-Ones in a block | | | | X | | | |
| Spectral | | | | X | | | |
| The Non-overlapping Template Matching | | | | X | | | |
| The Overlapping Template Matching | | | | X | | | |
| Maurer's "Universal Statistical" | | | | X | | | |
| The Linear Complexity | | | | X | | X | |
| The (Approximate) Entropy | | | | X | X | | |
| Cusums | | | | X | | | |
| The Random Excursions | | | | X | | | |
| The Random Excursions Variant | | | | X | | | |
| Chi-square | | | | | X | | |
| Arithmetic Mean | | | | | X | | |
| Monte Carlo Value for $\pi$ | | | | | X | | |
| Binary derivative | | | | | | X | |
| Change point | | | | | | X | |
| Sequence complexity | | | | | | X | |
| Equidistribution | | | | | | | X |
| Sum of Distributions | | | | | | | X |
| Ising Model | | | | | | | X |
| Random Walk (2D) | | | | | | | X |

Table 3.  Tests included in Kunth, Diehard, Dieharder, NIST, ENT and SPRNG batteries.

| Test / Battery | Small Crush (10 tests) | Crush (96 tests) | Big Crush (106 tests) | Rabbit (38 tests) | Alphabit (17 tests) | BlockAlphabit (17 tests) |
|---|---|---|---|---|---|---|
| Birthday Spacings | X | X | X | | | |
| Collision | X | X | X | | | |
| Coupon Collector | X | X | X | | | |
| Gap | X | X | X | | | |
| Hamming Independence | X | X | X | X | X | X |
| Maximum of t | X | X | X | | | |
| Random Walk One | X | X | X | X | X | X |
| Rank of a Binary Random Matrix | X | X | X | X | | |
| Simplified Poker | X | X | X | | | |
| Weighted Distribution | X | X | X | | | |
| Appearance Spacings | | X | X | X | | |
| Autocorrelation | | X | X | X | | |
| Close Pairs Bit Match | | X | | X | | |
| Closest Pairs | | X | X | | | |
| Collision-Permutation | | X | X | | | |
| Fourier 1 | | | | X | | |
| Fourier 3 | | X | X | X | | |
| GCD | | X | X | | | |
| Hamming Correlation | | X | X | X | X | X |
| Hamming Weights | | X | X | X | | |
| Lempel Ziv | | X | X | | | |
| Linear Complexity | | X | X | | | |
| Longest Head Run | | X | X | | | |
| Periods in Strings | | X | X | X | | |
| Permutation | | X | X | | | |
| Runs | | X | X | X | | |
| Runs of Bits | | X | X | | | |
| Sample Correlation | | X | X | | | |
| Sample Mean | | X | X | | | |
| Sample Product | | X | X | | | |
| Savir2 | | X | X | | | |
| Serial | | X | X | | | |
| Sumcollector | | X | X | | | |
| Multinomial Bits | | | | X | X | X |

Table 4.  Batteries and test included in TestU01.

| Battery | Advantages | Disadvantages |
|---|---|---|
| Knuth | First designed battery<br>Establishes the basis for the design of the others batteries | Objectively bad generators can pass the tests ($p-value \geq \alpha$)<br>Not currently used |
| Diehard | More rigorous tests than those presented by Knuth<br>It is more difficult that $p-value > \alpha$<br>It is more complete than the previous one | Fixed parameters<br>Limited sample size<br>Data must be in a binary file in the form of 32-bit integers (exactly) |
| Dieharder | More complete battery than the previous ones<br>Friendly interface<br>Open source<br>Can be used in R interface in Linux or Unix operating systems | Fixed parameters<br>Recommended for random numbers rather than bit sequences |
| NIST | More complete battery than the previous ones<br>Widely used<br>Designed for cryptographic apllications | Fixed parameters<br>Dependence between tests |
| TESTU01 | It includes most of the literature tests and implements the main RNGs<br>Flexibility to choose subsets of tests<br>Ability to work at a uniform number level as well as at a bit level | Depending on the selected subset it can take a considerable amount of computing time |
| ENT | Consists of five simple statistics, which are performed extremely quickly | Dependence between some of the obtained results |

Table 5.  Advantages and disadvantages of Knuth's, Diehard, Dieharder, NIST, Test U01 and ENT batteries.

| NIST TESTS | TESTING FUNCTION IN TESTU01 BATTERY |
|---|---|
| The Frequency (Monobit) test | sstring_HammingWeight2 with $L = n$ |
| Frequency test within a block | sstring_HammingWeight2 |
| The Runs test | sstring_Run |
| Tests for the Longest-Run-of-Ones in a block | sstring_LongestHeadRun |
| The Binary Matrix Rank test | smarsa_MatrixRank |
| The Discrete Fourier Transform (Spectral) test | sspectral_Fourier1 |
| The Non-overlapping Template Matching test | smarsa_CATBits |
| The Overlapping Template Matching test | smultin_MultinomialBitsOver (*) |
| Maurer's "Universal Statistical" test | svaria_AppearanceSpacings |
| The Linear Complexity test | scomp_LinearComp |
| The Serial test | smultin_MultinomialBitsOver with Delta = 1 |
| The Approximate Entropy test | smultin_MultinomialBitsOver with Delta = 0; sentrop_EntropyDiscOver; sentrop_EntropyDiscOver2 |
| The Cumulative Sums (Cusums) test | swalk_RandomWalk1 |
| The Random Excursions test | swalk_RandomWalk1 (*) |
| The Random Excursions Variant test | swalk_RandomWalk1 (*) based on the times that the excursion returns to 0. |

Table 6. NIST vs TestU01 batteries.